\documentclass[11pt]{article}
\usepackage{moriond,epsfig}

\def\mathfont#1{\ifmmode{#1}\else{$#1$}\fi} %for math font     
\def\lae{\mathrel{<\kern-1.0em\lower0.9ex\hbox{$\sim$}}}  
\def\gae{\mathrel{>\kern-1.0em\lower0.9ex\hbox{$\sim$}}}  
\def\ergsec{\mathfont{ {\rm ergs\ s}^{-1}}}

\def\be{\begin{equation}}
\def\ee{\end{equation}}
\def\bea{\begin{eqnarray}}
\def\eea{\end{eqnarray}}

\begin{document}
\noindent
To appear in ``Proceedings of XXI Moriond
conference: Galaxy Clusters and the High Redshift Universe Observed
in X-rays'', edited by D. Neumann, F. Durret, \& J. Tran Thanh Van
\vspace*{4cm}
\title{X-RAY STRUCTURE IN CLUSTER COOLING FLOWS AND ITS RELATIONSHIP
TO STAR FORMATION AND POWERFUL RADIO SOURCES}

\author{B.R. McNamara }

\address{Department of Physics \& Astronomy, Ohio University,\\
Athens, OH 45701, USA}

\maketitle\abstracts{Analyses of Chandra's first images of 
cooling flow clusters find  smaller cooling rates 
than previously thought.
Cooling may be occurring preferentially near regions of 
star formation in central cluster galaxies,
where the local cooling and star formation rates agree to within
factors of a few.
The radio sources in central cluster galaxies are interacting
with and are often displacing the hot, intracluster gas.
X-ray ``bubbles'' seen in Chandra images are used to 
measure the amount of energy radio sources deposit 
into their  surroundings,
and they may survive as fossil records of ancient radio activity.
The bubbles are vessels that transport
magnetic fields from giant black holes to the outskirts
of clusters.}

\section{Introduction}

The ``cooling flow problem,'' i.e., the 
fate of matter putatively
cooling below X-ray temperatures in the cores of clusters,
has stumped astronomers for over two decades (Fabian 1994).  This problem remains unsolved in part because 
the previous generation of X-ray telescopes was unable to resolve 
the cooling regions of clusters.    The Chandra X-ray
Observatory is poised to make great strides toward solving this problem
with its unique ability to 
resolve the inner several tens of kiloparsecs of clusters, where
the cooling time is shortest, and recently-accreted, 
$10-100$ K gas and young stars are observed.  Chandra's more than one
hundred-fold leap in combined spatial and spectral resolution is 
allowing astronomers
to measure and compare the X-ray cooling rates and optical star formation
rates on the same spatial scales.
In addition, Chandra is producing crisp images of 
the hot gas surrounding powerful radio sources in
central dominant cluster galaxies (CDGs) located at the bases of 
cooling flows.  This is a significant advance, because
radio sources can disturb and heat the surrounding hot gas, and 
possibly reduce the rate of cooling. The degree of heating and cooling
should be imprinted on the temperature and density structure of
the gas.  When viewed at high resolution, this structure can be compared
directly to local heat sources, such as the radio sources themselves and supernovae
associated with star formation.   As Chandra's first images of clusters 
arrived, it became immediately clear that the hot gas in cluster cores 
is in a complex state, and that Chandra would usher in an era of 
new and exciting cluster science.
Here, I briefly review new Chandra results on accretion-driven
star formation and interactions between radio plasma and hot gas 
in cooling flows as I understand them at this time.

\section{X-ray Cooling Rates \& Star Formation Rates}

The several Chandra studies of cooling flows reported 
thus far find cooling rates reduced by factors of 5 to 10 compared to
those derived from 
ROSAT data (Fabian et al. 2000a; McNamara et al 2000a,b; David et al. 2000).  
Spectroscopic evidence for multiphase, cooling gas is seen only in the inner
few tens of kpc of CDGs, where
the cooling time is $\lae 6\times 10^8$ yr.
Although inwardly decreasing gas temperature gradients
and cooling times less than 1 Gyr
are observed to cluster radii as large as a few tens of kpc, the
gas there can often be adequately described using single temperature
thermal models  (David et al. 2000).  How this gas 
avoids becoming multiphase and cooling to low temperatures is a new
challenge posed by the Chandra data (Fabian et al. 2000a).   

Although the cooling rates are more moderate
than previously reported, cool, multiphase gas (i.e., gas with a range
of temperature and density) is seen preferentially
in the central regions of CDGs.  Several groups have found strong correlations between 
X-ray cooling rate and the strength of
star formation in central cluster galaxies (Fig. 1, McNamara 1997, Cardiel et al. 1998).  The long-standing
problem has been that the X-ray cooling rates integrated over the
entire central $\sim 100$ kpc cooling radius exceeded 
the star formation rates by one to two orders of magnitude (assuming
the local initial mass function and a reasonable range of star 
formation histories). Furthermore,  star formation
is generally observed only in the inner few to few tens of kpc, and not
over the entire cooling region (Allen 1995; Cardiel et al. 1998, Crawford et al. 1999).  Multiphase gas is now seen by Chandra in several clusters
in dense clouds and filaments near the sites of star formation.
In these regions, the star formation and cooling rates are within factors of
$\sim 2$ of each other (McNamara et al. 2000b, Fabian this conference).  
The preliminary Chandra results bolster the interpretation of 
Fig. 1 that cooling flows are indeed fueling star formation.  
But the cooling rates are probably several to
several tens of solar masses per year, rather than several hundreds
of solar masses per year.  Although Chandra analyses have made
significant progress toward solving the cooling flow problem, 
not all of the putatively cooling material has been adequately 
accounted for, and the data do not by themselves prove that
cooling flows are inducing and fueling star formation.  
The influence of other
mechanisms, such as mergers or collisions with gas-rich galaxies,
must continue to be explored.  

%\subsection{Figures}\label{subsec:fig}
%{\em $\backslash$section$\ast$\{Acknowledgments\}}.

\begin{figure}
%\rule{5cm}{0.2mm}\hfill\rule{5cm}{0.2mm}
\hskip 3.5cm
\psfig{figure=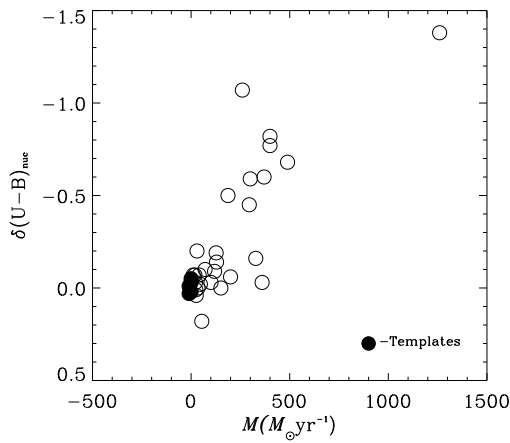,height=2.5in}
%\psfig{figure=test1.ps,height=2.0in}
%\rule{5cm}{0.2mm}\hfill\rule{5cm}{0.2mm}
\caption{Correlation between central $U-B$ continuum
color excess and total cooling rate.  The filled points represent
non-accreting galaxies with normal colors.
\label{fig:radish}}
\end{figure}

\section{Interactions Between Radio Sources and the keV Gas}

It is well known that radio sources interact
with and may energize the emission nebulae in cooling flows (Baum 1992),
and that they play a role in triggering star formation (see references
in McNamara 1999).  Chandra has since confirmed, in vivid detail, 
the early {\it ROSAT} evidence 
that radio sources are also interacting with the X-ray-emitting gas 
in clusters (e.g., B\"ohringer et al. 1993).  
In addition to showing that radio sources can have
a major impact on state of the hot gas surrounding them,  
these images are revealing the nature of radio
sources themselves.  For instance, we do not confidently understand
their ages, how long they persist,
their composition and kinetic energy content,
and how they evolve dynamically.  Before we can answer these questions,
we need a detailed picture of the physical conditions
of the medium in which radio sources move.  Chandra is giving 
us this picture.

\subsection{Cavities in the keV Gas} 

Chandra images of several clusters
with X-ray-bright cores have surface brightness depressions--cavities--in 
the hot, keV gas tens of kpc in size (Fig. 2).  Prominent examples are
seen in the Perseus cluster (Fabian et al. 2000b), Hydra A (McNamara et al.
2000a), Abell 2597 (McNamara et al 2000b, 2001), and Abell 2052 (Sarazin 2001).
In most cases the cavities are filled with radio emission
from the lobes of twin-jet radio sources whose luminosities exceed
$10^{42}~ \ergsec$. The gas along the rims of the cavities is close to
being the coolest gas emitting at keV temperatures in these clusters,
and no evidence for strong shocks immediately surrounding the
radio jets or lobes has been found.  The radio lobes appear 
to be gently displacing and are confined by the keV gas as they expand.  

\begin{figure}
%\rule{5cm}{0.2mm}\hfill\rule{5cm}{0.2mm}
\hskip 4.5cm
\psfig{figure=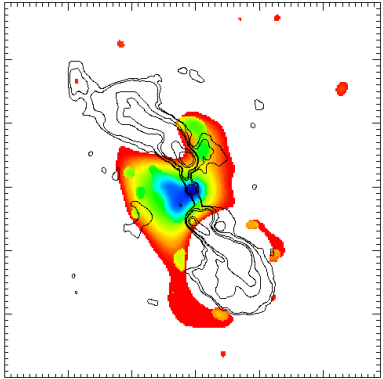,height=2.0in}
%\rule{5cm}{0.2mm}\hfill\rule{5cm}{0.2mm}
\caption{40 ksec exposure Chandra image of the central region of the 
Hydra A cluster (color). The X-ray image has been smoothed and 
filtered with wavelets in order to show the 
brightest emission. The 6 cm VLA Radio image of Hydra A
(contours) is superposed. See McNamara et al. (2000a) for details.
\label{fig:rad}}
\end{figure}

If the cavities, or bubbles, are supported by internal pressure 
from magnetic fields, cosmic rays, or a dilute, very hot plasma, 
they should rise outward into the ICM by buoyancy
(Churazov et al. 2000, McNamara et al. 2000).   The time required
to rise to distances of a few tens of kpc from the nucleus of the
host galaxy is $t_{\rm rise}\sim 10^{7.5-8}$ yr. If  this timescale
exceeds the on-time of the radio source, the radio surface brightness
in the cavities should fade as the energetic particles radiate away
their energy, leaving the long-term evolution of the bubble uncertain.

\subsection{Ghost Cavities and Radio Source Evolution}

CDGs in clusters such as Perseus (Fabian et al. 2000b) 
and Abell 2597 (McNamara et al 2000b, 2001) have ``ghost'' 
cavities in their X-ray emission that lie well beyond  the inner 
cavities associated with their nuclear radio sources.  These ghost cavities
are devoid of bright radio emission at wavelengths 
shorter than several centimeters.  If ghost cavities are all that
remains of a previous radio outburst that occurred $\sim t_{\rm rise}$ ago,
their existence in two clusters implies an evolutionary scenario
for powerful radio sources that can be reconciled with
the high incidence of radio activity observed in cooling flow CDGs.
More than 70\% of CDGs in clusters with dense, 
high surface brightness X-ray emission (cooling flows) harbor 
relatively powerful
radio sources, while less than 20\% of CDGs in non cooling flow 
clusters are radio-bright (Burns 1990).  
This trend implies that radio sources in cooling flows either persist
on Gyr timescales, or they recur with high frequency.  
Given what little we know of the demographics of
bubbles, their existence implies the recurring burst model
with a cycling time of $50-100$ Myr.

\subsection{Energizing and Magnetizing the keV Gas}

The sizes and pressures surrounding these bubbles, which are
directly measured from Chandra data, provide an estimate of the $PdV$ work
done by radio sources on the surrounding medium.  In Hydra A, 
the work is comparable to its radio luminosity for a
reasonable range of radio ages.  Therefore, there is little compelling
evidence for kinetic
luminosity that greatly exceeds its radio luminosity.  In Abell 2597, 
the energy associated with its ghost cavities is comparable
to the energy of the present-day radio source situated in close
proximity to the nucleus.  Therefore, in this case
the previous radio burst was probably similar to the one we see today.

The energy associated with bubbles is substantial, $\sim 10^{58-59}~{\rm erg}$.
If CDGs produce between $10-100$ bubbles over their lifetimes, 
the amount of energy they would deposit
into the ICM in the form of magnetic fields, cosmic rays, and 
heat is $\gae 10^{59-61}$ erg.  This energy would be comparable to the
the total thermal energy of the X-ray-emitting plasma in the
inner regions of clusters. Clusters are
magnetized  (Clarke, Kronberg, \& B\"ohringer 2001), 
and these bubbles may be vessels 
that transport magnetic fields from giant, central black 
holes to the outskirts of clusters.

\section*{Acknowledgments}

I would like to acknowledge 
my collaborators, in particular, Michael Wise, Paul Nulsen, Craig Sarazin,
and Larry David for their contributions to the work discussed here.

\end{document}